\documentclass[twocolumn,fleqn]{autart}

\usepackage{amsmath}
\usepackage{amssymb}
\usepackage{graphicx}
\usepackage{algorithm}
\usepackage{algpseudocode}
\usepackage{multicol}
\usepackage{multirow}
\usepackage{cite}
\usepackage{xcolor}
\usepackage{svg}
\usepackage{pdfpages}
\usepackage{epstopdf}
\usepackage{mathtools}
\usepackage{subfig}
\usepackage{caption}

\newtheorem{theorem}{Theorem}
\newtheorem{lemma}{Lemma}

\graphicspath{{images/}}

\begin{document}


\begin{frontmatter}

\title{Threshold Greedy Based Task Allocation for Multiple Robot Operations\thanksref{footnoteinfo}} 
\thanks[footnoteinfo]{\\ 
* Corresponding author. 
}

\author[Cranfield]{Teng Li}\ead{tengli@cranfield.ac.uk}, 
\author[Cranfield]{Hyo-Sang Shin*}\ead{h.shin@cranfield.ac.uk}, 
\author[Cranfield]{Antonios Tsourdos}\ead{a.tsourdos@cranfield.ac.uk}  

\address[Cranfield]{Cranfield University, Cranfield MK43 0AL, UK}

\begin{keyword}                           
Multi-robot systems, decentralised task allocation, submodular welfare maximisation, lazy decreasing threshold.                
\end{keyword}                             

\begin{abstract}                          
This paper deals with large-scale decentralised task allocation problems for multiple heterogeneous robots with monotone submodular objective functions. One of the significant challenges with the large-scale decentralised task allocation problem is the NP-hardness for computation and communication.  This paper proposes a decentralised Decreasing Threshold Task Allocation (DTTA) algorithm that enables parallel allocation by leveraging a decreasing threshold to handle the NP-hardness. Then DTTA is upgraded to a more practical version Lazy Decreasing Threshold Task Allocation (LDTTA) by combining a variant of Lazy strategy. DTTA and LDTTA can release both computational and communicating burden for multiple robots in a decentralised network while providing an optimality bound of solution quality. To examine the performance of the proposed algorithms, this paper models a multi-target surveillance scenario and conducts Monte-Carlo simulations. Simulation results reveal that the proposed algorithms achieve similar function values but consume much less running time and consensus steps compared with benchmark decentralised task allocation algorithms.
\end{abstract}

\end{frontmatter}

\section{Introduction}
\label{sec: Introduction}

Multi-Robot Systems (MRS) have been gaining increasing attention thanks to their ability to coordinate simultaneous or co-operate to achieve common goals. MRS provides some fundamental strengths that could not be achieved with single-agent systems, e.g., increased flexibility, enhanced reliability and resilience, simultaneous broad area coverage or capability to operate outside the communication range of base stations. Specific applications under consideration include search and rescue \cite{scherer2015autonomous, kurdi2016bio},  precision agriculture \cite{barrientos2011aerial, milics2019application}, and large area surveillance \cite{avellar2015multi, capitan2016cooperative, li2017potential, gu2018multiple}. 

Efficient cooperation of MRS is a vital part for their successful operations and effective assignment, termed as task allocation, of the available resources is the key enabler of such cooperation. This is because the strength of MRS hinges on the distributed nature of the sensing resources available, making the successful assignment of these resources key to maximising its operational advantages. 

The main issue with the task allocation problem is that it has been proven to be NP-hard in general \cite{shin2016uav}. This means that task allocation problems require exponential time to be solved optimally, thus require a careful craft of approximation strategies. A key trade-off that must be performed concerns the optimality of the solution versus the computational complexity. 

In terms of handling the NP-hardness of computation, extensive approaches have been developed, e.g., the heuristic approach and the approximation approach. Heuristic approach involves genetic algorithm \cite{bai2018integrated, jose2016task, zhu2018multi}, ant colony optimisation \cite{ilie2013multi, pendharkar2015ant, boveiri2016novel}, market-based algorithms \cite{gerkey2002sold, choi2009consensus, morgan2016swarm, lee2018resource, otte2019auctions} etc.. Heuristic algorithms can achieve feasible solutions with certain convergence speed. However, these algorithms cannot provide any approximation guarantee for the solution quality. Approximation approach solves task allocation problems efficiently while providing an optimality bound of solution quality \cite{ondracek2015solving, segui2015decentralised, qu2015distributed, kumar2017decentralized, williams2017decentralized, ding2017multi, gharesifard2017distributed, seo2018task, grimsman2018impact, corah2018distributed, qu2019distributed, corah2019distributed, zhou2019sensor, sun2019exploiting}, if the problem meets certain conditions, e.g., submodularity. Submodularity is a ubiquitous feature in the optimisation problems where the marginal gain of one element will diminish as more elements have already been selected \cite{balcan2015learning}. The detailed definition of submodularity will be given in Section \ref{sec: Preliminaries}. It is well known that the Sequential Greedy Algorithm (SGA) can provide an optimality bound of 1/2 for maximising monotone submodular objective functions subject to a partition matroid constraint \cite{nemhauser1978analysis}.

This paper focuses on task allocation algorithms that utilise the approximation approach and can be leveraged for decentralised task allocation. Note that in a centralised architecture, all the agents and environment data is communicated to a centralised entity. This may not be possible in some realistic scenarios because relying on a central entity could remove resilience or the bandwidth to communicate all the information may not be available. Capability to be decentralised can relax such issues and thus enabling decentralisation provides an option that could be beneficial for extending MRS operations in practice.  

The decentralised approximation approach has been frequently applied in the task allocation of MRS applications. The work \cite{qu2015distributed} presented a distributed SGA for a large group of earth-observing satellites to automatically assign themselves locations based on their local information and communication. Williams et al. \cite{williams2017decentralized} investigated the surveillance mission in an urban environment applying decentralised SGA with the consideration of the intersection of multiple matroid constraints. Sun et al. \cite{sun2019exploiting} solved the multi-agent coverage problems using distributed SGA in an environment with obstacles and provided a tighter optimality bound by analysing the curvature of submodularity. The decentralised approximation approach was also applied in search and localisation \cite{ding2017multi} and sensor network \cite{kumar2017decentralized, corah2018distributed} to get a task allocation solution efficiently.

The communicating complexity is another aspect of the NP-hardness in the decentralised task allocation problems. In decentralised networks, robots need to communicate with each other frequently and make a consensus before they can find an effective task allocation solution. Too many consensus steps could impact the efficiency of the task allocation process and put a heavy burden on the onboard communicating units.

In terms of relaxing the communicating complexity, several works of the decentralised approximation approach have been studied to reduce the reliance on the global information consensus. The benchmark decentralised task allocation algorithm CBBA \cite{choi2009consensus} can provide an undiminished optimality bound of 1/2 and reduce communicating complexity by building task bundles. These task bundles enable parallel task allocation hence reducing the number of global information consensus steps. However, as mentioned in \cite{choi2009consensus}, robots need to rebuild their bundles continually once any constraint conflict appears. The issue with the bundle rebuilding is that evaluating objective functions is usually time-consuming, which incurs observably increased computational complexity. Gharesifard et al. \cite{gharesifard2017distributed} provided lower bounds on the performance of the distributed SGA according to the network topology where each agent only had limited information of other agents' selections. Grimsman et al. \cite{grimsman2018impact} extended the work of Gharesifard et al. \cite{gharesifard2017distributed} and suggested the best system topology design. Although these two works reduced the communicating complexity of distributed SGA, the quality of solutions without global information consensus was also discounted. To handle this issue, Qu et al. \cite{qu2019distributed} proposed an assumption of admissible task sets for a group of observing satellites. With the help of this assumption, distributed SGA could achieve an undiminished optimality bound even though only local information and communication were accessible. The assumption is reasonable for homogeneous and spatially static agents, but difficult to be applied for general heterogeneous moving MRS. 

Previous works indicate that the goals of reducing computational complexity, reducing communicating complexity, and achieving good solution quality are usually conflicting in the large-scale task allocation problems, unless resorting to particular assumptions \cite{corah2019distributed, qu2019distributed}. One intuitive research question would be how to balance these goals in general large-scale task allocation problems.

This paper aims to develop decentralised algorithms that can reduce both computational and communicating complexity while retaining a good optimality bound of the solution for general heterogeneous MRS in large-scale task allocation problems. To this end, we propose an efficient decentralised task allocation algorithm, which is named as Decreasing Threshold Task Allocation (DTTA), by leveraging a decreasing threshold \cite{badanidiyuru2014fast, buchbinder2016comparing} to the decentralised SGA. 

In each iteration of the original decentralised SGA, each robot is required to evaluate all the remaining tasks then make a consensus with other robots to find a task-robot pair that can provide the largest marginal value. Hence, only one task can be allocated during each consensus step with SGA. The exploitation of the decreasing threshold concept relaxes this limitation. Hence, our proposition allows DTTA to consume fewer numbers of objective function evaluations and consensus steps and hence to reduce computational and communicating complexity.

This paper also extends the proposed DTTA algorithm to a more practical version, Lazy Decreasing Threshold Task Allocation (LDTTA), by combining a variant of the Lazy Greedy \cite{minoux1978accelerated}. In LDTTA, the remaining tasks are sorted in descending order according to their marginal values. During each iteration of LDTTA, we propose each robot to reevaluate the first element from the sorted remaining tasks and to compare it with the current threshold instead of the second element which is used in the original Lazy Greedy algorithm \cite{minoux1978accelerated}. The Lazy Greedy concept developed could further relax the computational and communicating complexity.

The proposed DTTA and LDTTA algorithms enable parallel allocation with the help of a decreasing threshold instead of building task bundles \cite{choi2009consensus} or making particular assumptions \cite{corah2019distributed, qu2019distributed}. To the best of our knowledge, this paper is the first work that can enable parallel allocation without incurring increased computational complexity or resorting to additional specific assumptions. 

Theoretical analysis reveals that DTTA and LDTTA achieve an optimality bound of at least ($1/2-\epsilon$) with computational complexity of $O(\frac{r}{\epsilon}\ln\frac{r}{\epsilon})$ for each robot, where $r=|\mathcal{T}|$ is the number of tasks. Here, $\epsilon$ is a threshold decreasing parameter for the trade-off of solution quality versus computational and communicating complexity. As expected, LDTTA outperforms DTTA, especially in terms of computational complexity.

The performances of DTTA and LDTTA are validated via numerical simulations. For the simulation-based validation, a scenario of multi-target surveillance mission using multiple heterogeneous UAVs is modelled. Monte-Carlo simulations are conducted to compare the performances of DTTA and LDTTA with those of SGA and CBBA. Simulation results demonstrate that the proposed algorithms achieve almost the same objective function values, but consume much less running time and consensus steps, compared with benchmark algorithms. The simulation results also confirm that the trade-off of solution quality, running time, and consensus steps can be obtained by adjusting the threshold decreasing parameter $\epsilon$.

The rest of the paper is organised as follows. Section \ref{sec: Preliminaries} provides basic concepts and definitions. The proposed algorithm is described in detail and analysed in Section \ref{sec: Algorithms and Analysis}. Section \ref{sec: Numerical Simulations} models a surveillance scenario and compares the performances of DTTA and LDTTA with those of benchmark algorithms through numerical simulations. A summary of contributions and future research directions are offered in  Section \ref{sec: Conlusions}.

\section{Preliminaries}
\label{sec: Preliminaries}

This section presents the necessary definitions and fundamental concepts that are related to the proposed decentralised task allocation algorithms.
    
\begin{defn}
\label{def: TAsk allocation}
(Task Allocation) The task allocation problem is to allocate a set of tasks $\mathcal{T}$ to  a set of robots $\mathcal{A}$ so as to maximise the total value measured as 
\begin{equation}
\label{def: TA}
f(\mathcal{T},\mathcal{A})=\sum_{a \in \mathcal{A}}f_a(\mathcal{T}_a),
\end{equation}
where  $\mathcal{T}_a$ is the subset of tasks allocated to robot $a$, $f_a:2^{\mathcal{N}}\rightarrow\mathbb{R}_{\geq 0}$ is the objective function for robot $a$. The notations introduced in this definition are also used in the following part of this paper.
\end{defn}

\begin{defn}{
\label{def:submodularity}
(Submodularity \cite{feldman2017greed}) A set function $f:2^{\mathcal{N}}\rightarrow\mathbb{R}$ is \textit{submodular}, if $\forall~X,Y\subseteq\mathcal{N}$,
\begin{equation*}
f(X)+f(Y)\geq f(X\cap Y)+f(X\cup Y),
\end{equation*}
where $\mathcal{N}$ is a finite set. Equivalently, $\forall~A\subseteq B \subseteq \mathcal{N}$ and $u\in\mathcal{N}\backslash B$,
\begin{equation}
\label{def:diminishing returns}
f(A\cup\{u\})-f(A)\geq f(B\cup\{u\})-f(B).
\end{equation}
}
\end{defn}

Eqn. (\ref{def:diminishing returns}) is known as the diminishing returns, which is an important property of submodular functions. Specifically, the marginal gain of a given element $u$ will never increase as more elements have already been selected.

\begin{defn}
\label{def:marginal_gain}
(Marginal gain \cite{krause2014submodular})  
For a set function $f:2^{\mathcal{N}}\rightarrow\mathbb{R}$, a set $S \subseteq \mathcal{N}$, and an element $u \in \mathcal{N}$, define the \textit{marginal gain} of $u$ given $S$ as
\begin{equation*}
\Delta f(u|S):=f(S \cup \{u\})-f(S).
\end{equation*}
\end{defn}

\begin{defn}
\label{def:monotonicity}
(Monotonicity \cite{krause2014submodular})  
A set function $f:2^{\mathcal{N}}\rightarrow\mathbb{R}$ is \textit{monotone}, if $\forall A \subseteq B \subseteq \mathcal{N}$, $f(A) \leq f(B)$. $f$ is \textit{non-monotone} if it is not monotone.
\end{defn}

This paper only considers monotone normalised (i.e. $f(\emptyset)=0$)  non-negative (i.e. $f(S) \geq 0$, $\forall S \subseteq \mathcal{N}$) submodular objective functions.

\begin{defn}
\label{def:matroid}
(Matroid \cite{badanidiyuru2014fast}) 
A matroid is a pair $\mathcal{M}=(\mathcal{N},\mathcal{I})$ where $\mathcal{N}$ is a finite set, and $\mathcal{I} \subseteq 2^\mathcal{N}$ is a collection of independent sets, satisfying:
\begin{itemize}
\item $\emptyset \in \mathcal{I}$ 
\item if $A \subseteq B, B \in \mathcal{I}$, then $A \in \mathcal{I}$ 
\item if $A, B \in \mathcal{I},|A|<|B|$, then $\exists~u \in B \backslash A ~\mbox{such that}  ~A \cup \{u\} \in \mathcal{I}$.
\end{itemize}
\end{defn}

Specifically, matroid constraints include uniform matroid constraint and partition matroid constraint. The uniform matroid constraint is also called cardinality constraint which is a special case of matroid constraint where any subset $S \subseteq \mathcal{N}$ satisfying $|S| \leq k$ is independent. Partition matroid constraint means that a subset $S$ can contain at most a certain number of elements from each of the disjoint partitions.

In this paper, one task can only be allocated to at most one robot, but one robot can take more than one task. Therefore, the task allocation problem can be solved as submodular maximisation subject to a partition matroid constraint, which is also known as Submodular Welfare Maximisation (SWM) \cite{vondrak2008optimal, segui2015decentralised}.

\section{Algorithms and Analysis}
\label{sec: Algorithms and Analysis}

This section describes the proposed task allocation algorithms and analyses their theoretical performances in details. First, the basic decentralised task allocation algorithm DTTA is introduced in Algorithm \ref{alg: decentralised DTTA}. Then, DTTA is extended to a more practical version (i.e., LDTTA) as presented in Algorithm \ref{alg: decentralised LDTTA} by developing and combining a variant of the Lazy Greedy strategy \cite{minoux1978accelerated}. Finally, the theoretical performances of the proposed algorithms are analysed through an equivalent reformed decreasing threshold greedy algorithm for submodular maximisation subject to a partition matroid constraint in Algorithm \ref{alg: Threshold Greedy Matroid}.

\subsection{Algorithms}     

Let us discuss the details of the DTTA and LDTTA algorithms. DTTA and LDTTA consist of two phases: initialisation phase (Algorithm \ref{alg: decentralised DTTA}, lines 1$\sim$4; Algorithm \ref{alg: decentralised LDTTA}, lines 1$\sim$10) and task assignment phase (Algorithm \ref{alg: decentralised DTTA}, lines 5$\sim$21; Algorithm \ref{alg: decentralised LDTTA}, lines 11$\sim$35). $\mathcal{N}_a$ is the set that contains the remaining tasks for robot $a$. $A$ is an auxiliary set that contains the ids of robots who are assigned tasks during each iteration. $A=\emptyset$ means that no robot can find a qualified task under the current threshold, which is a trigger of decreasing the threshold (Algorithm \ref{alg: decentralised DTTA}, line 6; Algorithm \ref{alg: decentralised LDTTA}, line 12). A qualified task is a task whose marginal value for robot $a$ given the selected task list $\mathcal{T}_a$ is no less than the current threshold $\theta$. $J$ is an auxiliary set that contains the corresponding task ids that are assigned to robots in $A$ during each iteration. $\mathcal{W}_a$ is an auxiliary set that contains the marginal values of tasks from $\mathcal{N}_a$ in Algorithm \ref{alg: decentralised LDTTA}.

In the initialisation phase, robot $a$ conducts preparatory work. Robot $a$ evaluates all tasks from $\mathcal{N}_a$ and finds the largest marginal value $\omega_a^{max}$ (Algorithm \ref{alg: decentralised DTTA}, line 2; Algorithm \ref{alg: decentralised LDTTA}, lines 2$\sim$8). After getting its $\omega_a^{max}$, robot $a$ makes a consensus with other robots through the $MaxCons$ function to find the globally largest marginal value $\omega_{a^*}^{max}$ (Algorithm \ref{alg: decentralised DTTA}, line 3; Algorithm \ref{alg: decentralised LDTTA}, line 9). The value of $\omega_{a^*}^{max}$ will be set as the initial threshold value $d$ (Algorithm \ref{alg: decentralised DTTA}, line 4; Algorithm \ref{alg: decentralised LDTTA}, line 10). 

In the task assignment phase, robot $a$ searches qualified tasks then negotiates with other robots to assign tasks and solve conflicts. First, robot $a$ tries to find one qualified task from the remaining tasks in $\mathcal{N}_a$. The notations $j_a^*$ and $\omega_a^*$ are used to store the qualified task and its marginal value (Algorithm \ref{alg: decentralised DTTA}, lines 7$\sim$9; Algorithm \ref{alg: decentralised LDTTA}, lines 14$\sim$17). Robot $a$ will stop searching once it finds one qualified task which is not necessarily the best one. There is no need to reevaluate all the remaining tasks in $\mathcal{N}_a$. Note that, in the decentralised SGA, each robot needs to reevaluate all the remaining tasks to find the task that can provide the largest marginal value. If robot $a$ cannot find a qualified task, the values of $\omega_a^*$ and $j_a^*$ will be set as 0 and none, respectively (Algorithm \ref{alg: decentralised DTTA}, lines 11$\sim$12; Algorithm \ref{alg: decentralised LDTTA}, lines 28$\sim$29). Then, robot $a$ coordinates with other robots through the $MaxCoor$ function to assign tasks and solve conflicts (Algorithm \ref{alg: decentralised DTTA}, lines 14$\sim$18; Algorithm \ref{alg: decentralised LDTTA}, lines 18$\sim$23, lines 30$\sim$32). If no robot can provide a qualified task, then the auxiliary set $A$ will be empty, and all robots will reduce the threshold to the next value $\theta(1-\epsilon)$ and reevaluate the remaining tasks until reaching the stop condition (Algorithm \ref{alg: decentralised DTTA}, line 5; Algorithm \ref{alg: decentralised LDTTA}, line 11).

\begin{algorithm}[]
\caption{decentralised DTTA for Robot $a$}
\textbf{Input:} $f_a:2^\mathcal{T}\rightarrow\mathbb{R}_{\geq0}, \mathcal{T}, \mathcal{A}, \epsilon$   \\
\textbf{Output:} A set $\mathcal{T}_a \subseteq \mathcal{T}$

\label{alg: decentralised DTTA}
\begin{algorithmic}[1]
\State $\mathcal{T}_a \leftarrow \emptyset$, $\mathcal{N}_a \leftarrow \mathcal{T}$
\State $\omega_a^{max} \leftarrow \max \limits_{j_a \in \mathcal{N}_a} f_a(j_a|\mathcal{T}_a)$
\State $a^*, \omega_{a^*}^{max} \leftarrow MaxCons(a, \omega_a^{max})$
\State $d \leftarrow \omega_{a^*}^{max}, r \leftarrow |\mathcal{T}|$, $A \leftarrow \{a^*\}$
\For {($ \theta = d; \theta \geq \frac{\epsilon}{r}d; \theta \leftarrow \theta(1-\epsilon))$}
    \While{$A \neq \emptyset$}
        \If{$\exists~j_a \in \mathcal{N}_a$ such that $\Delta f_a(j_a|\mathcal{T}_a) \geq \theta$}
            \State $\omega_a^* \leftarrow \Delta f_a(j_a|\mathcal{T}_a)$
            \State $j_a^* \leftarrow j_a$ 
        \Else
            \State $\omega_a^* \leftarrow 0$
            \State $j_a^* \leftarrow none$
        \EndIf
        \State $A, J \leftarrow MaxCoor(a, j_a^*, \omega_a^*)$
        \If{$a \in A$}
            \State $\mathcal{T}_a \leftarrow \mathcal{T}_a \cup \{j_a^*\}$
        \EndIf
        \State $\mathcal{N}_a \leftarrow \mathcal{N}_a \backslash J$
    \EndWhile
\EndFor \\
\Return $\mathcal{T}_a$
\end{algorithmic}
\end{algorithm}

The function $MaxCons$ and $MaxCoor$ in Algorithms \ref{alg: decentralised DTTA} and \ref{alg: decentralised LDTTA} represent the communication and negotiation among all robots. These two functions are described in details as follows.

$\bullet$ $MaxCons$ is the Maximum Consensus function. Robot $a$ sends its locally best marginal value $\omega_a^{max}$ together with the corresponding robot id $a$ to all its neighbouring robots who have direct communication connections with robot $a$. At the same time, robot $a$ also receives the same kind of information from its neighbouring robots. After receiving the information from its neighbouring robots, robot $a$ sends the updated information to all its neighbouring robots again until no information is updated. In this case, robot $a$ can pass the information for its neighbouring robots who have no direct connection between each other. All robots can make a consensus and finally find the globally largest marginal value $\omega_{a^*}^{max}$. It is proven that the Maximum Consensus function can reach the convergence in finite time \cite{giannini2016asynchronous, iutzeler2012analysis, olfati2004consensus, cortes2008distributed}. Note that this function is only utilised for one time by robot $a$ in the initialisation phase of the proposed algorithms.

$\bullet$ $MaxCoor$ is defined as the Maximum Coordination function. Robot $a$ sends the selected task id $j_a^*$ and its marginal value $\omega_a^*$ to all neighbouring robots and receives such information from its neighbouring robots. Similar to the function $MaxCons$, $MaxCoor$ also conducts a consensus on the information of all robots. Then, $MaxCoor$ returns the auxiliary robot set $A$ and auxiliary task set $J$ according to the following criteria: if there is no conflict between robots (i.e., no task is selected by more than one robot), then every robot and its qualified task will be added into $A$ and $J$, respectively; if there is any conflict (i.e., more than one robot selects the same task $j_a^*$), then the robot who has the largest marginal value for the task $j_a^*$ will be added into $A$, and the task $j_a^*$ will be added into $J$; if the input variable $\omega_a^*$ from all robots are zeroes, then both $A$ and $J$ will be empty which triggers the decrease of the threshold $\theta$. With this strategy, more than one task could be allocated during one iteration, which can reduce the number of task assignment iterations and consensus steps.

\begin{algorithm}[t]
\caption{decentralised LDTTA for Robot $a$}
\textbf{Input:} $f_a: 2^\mathcal{T}\rightarrow\mathbb{R}_{\geq0}, \mathcal{T}, \mathcal{A}, \epsilon$   \\
\textbf{Output:} A set $\mathcal{T}_a \subseteq \mathcal{T}$

\label{alg: decentralised LDTTA}
\begin{algorithmic}[1]
\State $\mathcal{T}_a \leftarrow \emptyset$, $\mathcal{N}_a \leftarrow \mathcal{T}$, $\mathcal{W}_a \leftarrow \emptyset$
\For{$j_a \in \mathcal{N}_a$}
    \State $\omega_{aj} \leftarrow \Delta f_a(j_a|\mathcal{T}_a)$
    \State $\mathcal{W}_a \leftarrow \mathcal{W}_a \cup \{\omega_{aj}\}$
\EndFor
\State Sort $\mathcal{W}_a$ in descending order
\State Sort $\mathcal{N}_a$ according to sorted $\mathcal{W}_a$
\State $\omega_a^{max} \leftarrow \omega_{a1}$
\State $a^*, \omega_{a^*}^{max} \leftarrow MaxCons(a, \omega_a^{max})$
\State $d \leftarrow \omega_{a^*}^{max}, r \leftarrow |\mathcal{T}|$, $A \leftarrow \{a^*\}$
\For {($ \theta = d; \theta \geq \frac{\epsilon}{r}d; \theta \leftarrow \theta(1-\epsilon))$}
    \While{$A \neq \emptyset$}
        \While{$\omega_{a1} \geq \theta$}
            \State $\omega_{a1} \leftarrow \Delta f_a(j_{a1}|\mathcal{T}_a)$
            \If{$\omega_{a1} \geq \theta$}
                \State $\omega_a^* \leftarrow \omega_{a1}$
                \State $j_a^* \leftarrow j_{a1}$
                \State $A, J \leftarrow MaxCoor(a, j_a^*, \omega_a^*)$
                \If{$a \in A$}
                    \State $\mathcal{T}_a \leftarrow \mathcal{T}_a \cup \{j_a^*\}$
                \EndIf
                \State $\mathcal{N}_a \leftarrow \mathcal{N}_a \backslash J$
                \State $\mathcal{W}_a \leftarrow \mathcal{W}_a  \backslash \mathcal{W}_a(J)$
            \Else
                \State Re-sort $\mathcal{W}_a$ and $\mathcal{N}_a$
            \EndIf
        \EndWhile
        \State $\omega_a^* \leftarrow 0$
        \State $j_a^* \leftarrow none$
        \State $A, J \leftarrow MaxCoor(a, j_a^*, \omega_a^*)$
        \State $\mathcal{N}_a \leftarrow \mathcal{N}_a \backslash J$
        \State $\mathcal{W}_a \leftarrow \mathcal{W}_a  \backslash \mathcal{W}_a(J)$        
    \EndWhile
\EndFor \\
\Return $\mathcal{T}_a$
\end{algorithmic}
// $\omega_{a1}$ is the first element of the sorted marginal value set $\mathcal{W}_a$; $j_{a1}$ is the first element of the sorted task set $\mathcal{N}_a$; $\mathcal{W}_a(J)$ is the set of marginal values that are corresponding to the tasks in $J$
\end{algorithm}

In the original Lazy greedy \cite{minoux1978accelerated}, the updated marginal value of the first element from the sorted remaining element list is compared with the formerly second marginal value. Instead, the updated first marginal value is compared with the current threshold in the variant of lazy greedy. The proposed new lazy greedy strategy is also applicable for the general decreasing threshold greedy algorithm \cite{badanidiyuru2014fast}.

Note that algorithm \ref{alg: decentralised LDTTA} could help robot $a$ to find qualified tasks in a more efficient way than Algorithm \ref{alg: decentralised DTTA} in practice. In the initialisation phase of Algorithm \ref{alg: decentralised LDTTA}, robot $a$ evaluates the tasks in $\mathcal{N}_a$ and sorts them in descending order according to their marginal values (Algorithm \ref{alg: decentralised LDTTA}, lines 2$\sim$7). During the task assignment phase of Algorithm \ref{alg: decentralised LDTTA}, robot $a$ checks whether the first marginal value $\omega_{a1}$ from the sorted marginal value set $\mathcal{W}_a$ is no less than the current threshold $\theta$ (Algorithm \ref{alg: decentralised LDTTA}, line 13). If $\omega_{a1} < \theta$, then there is no need for robot $a$ to continue searching for qualified tasks under the current threshold because submodularity guarantees that the marginal value of a task will never increase. Otherwise, robot $a$ reevaluates the first task $j_{a1}$ from the sorted task list $\mathcal{N}_a$ given the current allocated task set $\mathcal{T}_a$, and updates the first marginal value $\omega_{a1}$ from the sorted marginal value set $\mathcal{W}_a$ (Algorithm \ref{alg: decentralised LDTTA}, line 14). If this updated $\omega_{a1}$ is still no less than $\theta$, it implies that the first task $j_{a1}$ from the sorted task list $\mathcal{N}_a$ is a qualified task for robot $a$. Robot $a$ will use this task and its marginal value to negotiate with other robots (Algorithm \ref{alg: decentralised LDTTA}, lines 15$\sim$23). Otherwise, robot $a$ re-sorts $\mathcal{W}_a$ and $\mathcal{N}_a$ in decreasing order (Algorithm \ref{alg: decentralised LDTTA}, lines 24$\sim$26). If robot $a$ is unable to find a qualified task, it still needs to communicate with other robots and remove the tasks from $\mathcal{N}_a$ that are selected by other robots to prevent potential conflicts (Algorithm \ref{alg: decentralised LDTTA}, lines 28$\sim$32).

\subsection{Analysis}    

Algorithms \ref{alg: decentralised DTTA} and \ref{alg: decentralised LDTTA} have the same theoretical performance in terms of both optimality bound and computational complexity. In each iteration, both algorithms help robot $a$ to find a qualified task whose marginal value is no less than the current threshold $\theta$. Therefore, DTTA and LDTTA achieve the same theoretical optimality bound. In LDTTA, the remaining tasks from $\mathcal{N}_a$ are sorted in descending order according to their marginal values. Robot $a$ is more likely to find a qualified task from the front positions of $\mathcal{N}_a$, requiring less number of objective function evaluations. This is the reason why LDTTA could be more efficient than DTTA in practice. However, in the worst case, DTTA and LDTTA need to evaluate all the remaining tasks to find a qualified task. Therefore, DTTA and LDTTA have the same computational complexity in theory.

\begin{algorithm}[t]
\caption{equivalent Decreasing Threshold Greedy subject to matroid constraints}
\textbf{Input:} $f:2^\mathcal{N}\rightarrow\mathbb{R}_{\geq0}, \mathcal{N}, \mathcal{I}, r, \epsilon$   \\
\textbf{Output:} A set $S\in\mathcal{I}$
\label{alg: Threshold Greedy Matroid}
\begin{algorithmic}[1]
\State $S \leftarrow \emptyset$, $R \leftarrow \mathcal{N}$, $Q \leftarrow OPT$
\State $d \leftarrow \max \limits_{u \in \mathcal{N}}f(u)$

\For {($ \theta = d; \theta \geq \frac{\epsilon}{r}d; \theta \leftarrow \theta(1-\epsilon))$}
    \For {$u \in R$}
        \If {$S\cup\{u\} \notin \mathcal{I}$}
            \State $R \leftarrow R \backslash \{u\}$
        \Else
            \If {$\Delta f(u|S) \geq \theta$}
                \State $c \leftarrow u$
                \State $S_c \leftarrow S$
                \State $S \leftarrow S \cup \{c\}$
                \State $Q \leftarrow Q \cup \{c\}$
                \State Let $K_c\subseteq Q \backslash S$ be the smallest set such that $Q \backslash K_c \in \mathcal{I}$
                \State $Q \leftarrow Q \backslash K_c$
                \State $R \leftarrow R \backslash \{c\}$
            \Else
                \If{$\Delta f(u|S) < \frac{\epsilon}{r}d$}
                    \State $R \leftarrow R \backslash \{u\}$
                \EndIf
            \EndIf
        \EndIf
    \EndFor
\EndFor \\
\Return $S$
\end{algorithmic}
\end{algorithm}

For the convenience of analysing the theoretical performance, DTTA and LDTTA are transformed into an equivalent centralised version, i.e., Algorithm \ref{alg: Threshold Greedy Matroid}. We consider the ground set as a set of task-agent pairs ($\mathcal{N}:= \mathcal{T} \times \mathcal{A}$) and each task-agent pair as an element of the ground set ($u_{j,a}:=j \times a ~ \forall j \in \mathcal{T}, a \in \mathcal{A}$). Then, only one task-agent pair can be selected from the task-agent pairs that are corresponding to the same specific task. All the combinations of conflict-free task-agent pairs constitute $\mathcal{I}$, which is denoted as the collection of all independent sets. Additionally, according to the definition of Task Allocation, the objective function of one agent has no impact on the objective function of another agent as long as their selected task sets are disjoint. Therefore, the task allocation problem can be considered as SWM. We adapt the analysing strategy from \cite{feldman2017greed} and transform Algorithms \ref{alg: decentralised DTTA} and \ref{alg: decentralised LDTTA} to Algorithm \ref{alg: Threshold Greedy Matroid}.

In Algorithm \ref{alg: Threshold Greedy Matroid}, $S_c$, $Q$ and $K_c$ have no impact on the final solution $S$. They appear only for analysis: $S_c$ is a set containing the elements that have already been selected before the new element $c$ is added into $S$; $Q$ is a set that starts as the optimal solution $OPT$ and changes over time meanwhile keeps containing $S$; $K_c$ is a set that contains the element to be removed from $Q$ in order to keep $Q$ independent.

The theoretical performance of the proposed task allocation algorithms DTTA and LDTTA is summarised in Theorem \ref{thm: main_theorem}.

\begin{theorem}
\label{thm: main_theorem}
Both DTTA and LDTTA achieve an optimality bound of at least ($\frac{1}{2}-\epsilon$) for maximising monotone submodular objective functions. The computational complexity for each robot is $O(\frac{r}{\epsilon}\ln\frac{r}{\epsilon})$, where $r=|\mathcal{T}|$ is the number of all tasks, $\epsilon$ is the threshold decreasing parameter.
\end{theorem}

The computational complexity can be easily proven as follows.
\begin{pf}
Assume that there are totally $x$ number of iterations, then
\begin{equation}
(1-\epsilon)^x = \frac{\epsilon}{r}.
\end{equation}
Solving the above equation yields
\begin{equation}
x = \frac{\ln\frac{r}{\epsilon}}{\ln\frac{1}{1-\epsilon}} \leq \frac{1}{\epsilon}\ln\frac{r}{\epsilon}.
\end{equation}
For each robot, there are at most $r$ number of tasks to be evaluated during each iteration. Therefore, the computational complexity for each robot is $O(\frac{r}{\epsilon}\ln\frac{r}{\epsilon})$.
\end{pf}

In the following, the optimality bound of DTTA and LDTTA is analysed through Algorithm \ref{alg: Threshold Greedy Matroid} since they have precisely the same theoretical approximation performance. 

\begin{lemma}
\label{thm: lemma_1}
$f(S) > \frac{1}{1+\epsilon}f(Q)$.
\end{lemma}

\begin{pf}
At the end of each iteration in Algorithm \ref{alg: Threshold Greedy Matroid}, $Q$ is always independent and contains $S$, i.e. $S \subseteq Q \in \mathcal{I}$. The property of independent systems implies that $S \cup \{q\} \in \mathcal{I} ~ \forall q \in Q \backslash S$.  At the termination of Algorithm \ref{alg: Threshold Greedy Matroid}, $\Delta f(q|S) < \frac{\epsilon}{r}d ~ \forall q \in Q \backslash S$ and $f(S) \geq d$. Clearly, $|Q \backslash S| \leq r$, thus
\begin{align*}
\label{termination}
\sum\limits_{q \in Q \backslash S}\Delta f(q|S) &< \sum\limits_{q \in Q \backslash S}\frac{\epsilon}{r}d \\ 
&\leq \epsilon \cdot \frac{|Q \backslash S|}{r} f(S) \\
&\leq \epsilon \cdot f(S).
\end{align*}


Let $Q \backslash S = \{q_1, q_2, \cdots, q_{|Q \backslash S|}\}$, then
\begin{flalign*}
f(Q) - f(S) &= \sum \limits_{i=1}^{|Q \backslash S|}\Delta f(q_i|S \cup \{q_1, \cdots, q_{i-1}\}) \\
& \leq \sum\limits_{i=1}^{|Q \backslash S|}\Delta f(q_i|S) \tag{submodularity}\\
& = \sum\limits_{q \in Q \backslash S}\Delta f(q|S) \\
& < \epsilon \cdot f(S).
\end{flalign*}
Rearranging the above inequation yields 
\begin{equation*}
f(S) > \frac{1}{1+\epsilon}f(Q). 
\end{equation*}
\end{pf}

Lemma \ref{thm: lemma_1} indicates that, at the termination of Algorithm \ref{alg: Threshold Greedy Matroid}, $f(S)$ gets a close value to $f(Q)$ if the value of $\epsilon$ is small enough. This implies the elements in $Q \backslash S$, whose marginal values are less than $\frac{\epsilon}{r}d$, have very limited contribution to $f(S)$. Likewise, if the marginal value of a task for a robot is less than $\frac{\epsilon}{r}d$, then we consider this task negligible for this robot. This is the reason why we choose $\frac{\epsilon}{r}d$ as the terminal threshold. The task allocation algorithm terminates when all marginal values of the remaining tasks are less than $\frac{\epsilon}{r}d$.

\begin{lemma}
\label{thm: lemma_2}
$f(Q) > f(OPT) - \frac{1}{1-\epsilon}f(S)$.
\end{lemma}

\begin{pf}
According to the property of matroid, $K_c$ contains at most one element. Therefore,
\begin{equation}
\label{abs_kc}
|K_c \backslash S| \leq 1. 
\end{equation}
In the iteration where the element $c$ is selected, it implies that the marginal value of $c$ is no less than the current threshold $\theta$, i.e.
\begin{equation}
\label{threshold greater} 
\Delta f(c|S_c) \geq \theta. 
\end{equation}
If an element  $q \in K_c \backslash S$ was not selected before this iteration, then it implies that 
\begin{equation}
\label{threshold less} 
\Delta f(q|S_c) < \theta / (1-\epsilon).
\end{equation}
Combining Eqns. (\ref{threshold greater}) and (\ref{threshold less}), we get
\begin{equation} 
\label{c vs o}   
\Delta f(c|S_c) > (1-\epsilon) \Delta f(q| S_c) ~ \forall q \in K_c \backslash S.
\end{equation}
Additionally, in the iteration when $K_c \neq \emptyset$, the element contained in $K_c$ in this iteration will never appear in the $K_c$ again in other iterations. This implies that both $\{K_c\}_{c \in S}$ and  $\{K_c \backslash S\}_{c \in S}$ are disjoint.
According to the evolution of $Q$, the set $Q$ can be rewritten as
\begin{equation}
\label{definition of O}
Q = (OPT \backslash \cup_{c \in S}K_c) \cup S = (S \cup OPT) \backslash \cup_{c \in S}(K_c \backslash S).
\end{equation}
Denote $S$ as $\{c_1, c_2, \cdots, c_{|S|}\}$. It is clear that $S_{c_i} \subseteq S \subseteq (S \cup OPT) \backslash \cup_{c \in S}(K_c \backslash S)$. Using Eqn. (\ref{definition of O}), we have
\begin{flalign*}
& f(Q) = f((S \cup OPT) \backslash \cup_{c \in S}(K_c \backslash S)) && \\
&= f(S \cup OPT) &&\\
&\quad - \Delta f(\cup_{c \in S}(K_c \backslash S)|(S \cup OPT)\backslash \cup_{c \in S} (K_c \backslash S)) &&\\
&= f(S \cup OPT) &&\\
&\quad - \sum \limits_{i=1}^{|S|} \Delta f((K_{c_i} \backslash S)|(S \cup OPT)\backslash \cup_{1\leq j \leq i} (K_{c_j} \backslash S)) &&\\
&\geq f(OPT) &&\\
&\quad - \sum \limits_{i=1}^{|S|} \Delta f((K_{c_i} \backslash S)|(S \cup OPT)\backslash \cup_{1\leq j \leq i} (K_{c_j} \backslash S)) &&\\
&\geq  f(OPT) - \sum \limits_{i=1}^{|S|} \Delta f((K_{c_i} \backslash S)|S_{c_i}) \tag{submodularity} &&\\
&\geq f(OPT) - \sum \limits_{c \in S} \sum \limits_{q \in K_c \backslash S}\Delta f(q| S_c) \tag{submodularity} &&\\
&>  f(OPT) - \sum \limits_{c \in S} \sum \limits_{q \in K_c \backslash S}\frac{1}{1-\epsilon}\Delta f(c| S_c) \tag{Eqn. (\ref{c vs o}) } &&\\
&=  f(OPT) - |K_c \backslash S| \cdot \frac{1}{1-\epsilon} \cdot f(S) &&\\
&\geq f(OPT) - \frac{1}{1-\epsilon} \cdot f(S). \tag{Eqn. (\ref{abs_kc}) } &&
\end{flalign*}
\end{pf}

The proof of Theorem \ref{thm: main_theorem} is completed by combining Lemma \ref{thm: lemma_1} and Lemma \ref{thm: lemma_2},
\begin{equation*}
f(S) > \frac{1-\epsilon}{2-\epsilon^2} \cdot f(OPT) > (\frac{1}{2}-\epsilon)f(OPT).
\end{equation*}

According to the analytical results, the trade-off between the approximation ratio and computational complexity can be obtained by adjusting the threshold decreasing parameter $\epsilon$.
    

\section{Numerical Simulations}
\label{sec: Numerical Simulations}

This section verifies the proposed algorithms through Monte-Carlo simulations. The simulation scenario is modelled as a multi-target surveillance mission using a group of heterogeneous UAVs.

\subsection{Surveillance Scenario Modelling}
For validation, we develop a simple model of the surveillance scenario. In the model, it is assumed that there are a set of heterogeneous tasks ($\mathcal{T}$) that are randomly located on a $L \times L$ 2-D space. A set of heterogeneous UAVs ($\mathcal{A}$) equipped with different sensors is sent to carry out these tasks automatically. The task allocation mission aims to maximise the overall objective function value while using as less running time and consensus steps as possible. The constraint is that one task can only be allocated to one UAV, but one UAV can carry out multiple tasks.

The objective function of the surveillance mission for UAV $a$ is modelled as:
\begin{equation}
\label{eqn: f_a}
f_a(\mathcal{T}_a)=\sum_{j=1}^{|\mathcal{T}_a|}m_{aj} v_j \lambda_d^{\tau(\mathbf{p}_a^j)} \lambda_n^{\sigma(\mathbf{p}_a^j)}.
\end{equation}
The overall objective function of the task allocation for the surveillance mission can be obtained by combining Eqns. (\ref{def: TA}) and (\ref{eqn: f_a}):
\begin{equation}
\label{eqn: f_final}
f(\mathcal{T},\mathcal{A})=\sum_{a=1}^{|\mathcal{A}|}\sum_{j=1}^{|\mathcal{T}_a|}m_{aj} v_j \lambda_d^{\tau(\mathbf{p}_a^j)} \lambda_n^{\sigma(\mathbf{p}_a^j)}
\end{equation}
where $\mathcal{T}_a \subseteq \mathcal{T}$ is the task list containing all the tasks that are selected by UAV $a$ in sequence, $\mathbf{p}_a$ is the path of UAV $a$ generated according to the order that the tasks appear in $\mathcal{T}_a$, $\mathbf{p}_a^j$ represents the part of $\mathbf{p}_a$ from the initial position of UAV $a$ to the position of task $j$, $\tau(\mathbf{p}_a^j)$ is the length of the path fragment $\mathbf{p}_a^j$, and $\sigma(\mathbf{p}_a^j)$ is the number of tasks within the path fragment $\mathbf{p}_a^j$. 

Four factors are considered in the objective function of the surveillance mission.

$\bullet$ \textit{Task importance factor $v_j$:}
Different tasks have different values that are marked with an importance factor $v_j \in (0,1]$. UAVs tend to give priority to carrying out the tasks that are more valuable than others. 

$\bullet$ \textit{Task-UAV fitness factor $m_{aj}$:}
Tasks with different properties require different sensors to be detected effectively. Therefore, the fitness factor $m_{aj} \in (0,1]$ reflects the match fitness between the task $j$ and UAV $a$.

$\bullet$ \textit{Distance discount factor $\lambda_d$:}
The mission should be completed as soon as possible. With the help of the distance discount factor $\lambda_d \in (0,1]$, UAVs intend to carry out the nearest tasks to them firstly, and the lengths of all UAVs' paths should be balanced.

$\bullet$ \textit{Task number discount factor $\lambda_n$:}
It is usually risky to allocate a large number of tasks to one UAV, but only a few to others. The task number discount factor $\lambda_n \in (0,1]$ helps to balance the numbers of allocated tasks among all UAVs.

Note that, Eqn. (\ref{eqn: f_final}) is proven to be monotone, non-negative, and submodular in the appendix.

\subsection{Simulation Results}

The performances of the proposed algorithms are compared with the benchmark decentralised task allocation algorithms SGA and CBBA. We run 100 rounds of Monte-Carlo simulations then get the mean values of objective function value, running time, and consensus steps, respectively. The running time is measured as the number of objective function evaluations, which is independent on the computer status. The running time for each UAV approximately equals to the total running time divided by the number of UAVs. The performances of these algorithms are also compared through ratios with SGA as a baseline.

In the simulations, assume that there are totally 200 tasks to be carried out by a fleet of varying numbers of UAVs from 10 to 50 denoted as $N_a$, and let $L = 10 km$. Set the importance factor of each task as a uniformly random number $v_j \in [0.6,1.0]$, and the match fitness factor of each task-UAV pair as a uniformly random number $m_{aj} \in [0.5,1.0]$. Set the distance discount factor $\lambda_d=0.95$, and the task number discount factor $\lambda_n=0.98$, respectively. In algorithms DTTA and LDTTA, set the threshold decreasing parameter as $\epsilon=0.05$.

\begin{figure*}[ht]
\subfloat[Function Value]{\includegraphics[width=0.46\linewidth]{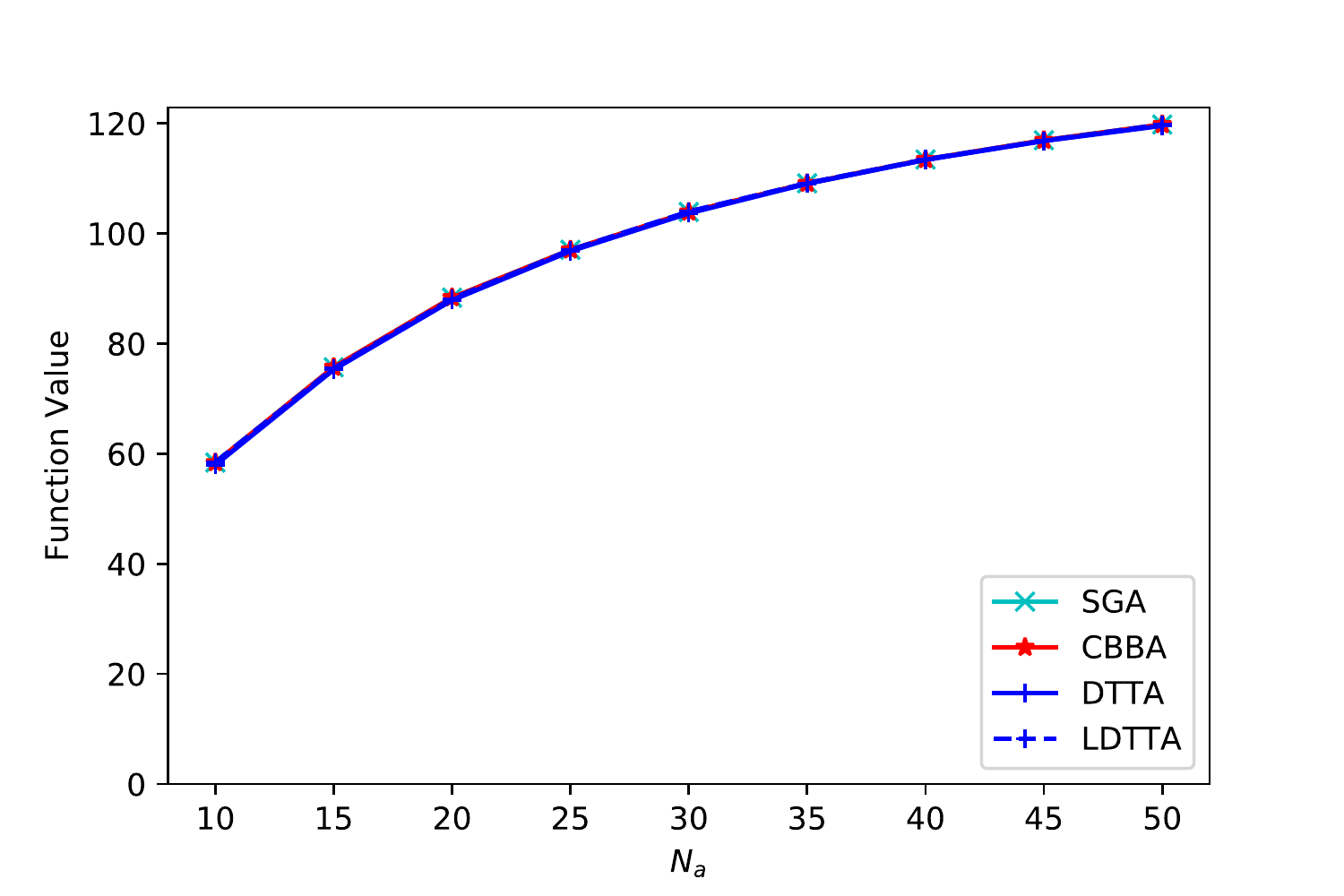}}\hfill
\subfloat[Running Time]{\includegraphics[width=0.46\textwidth]{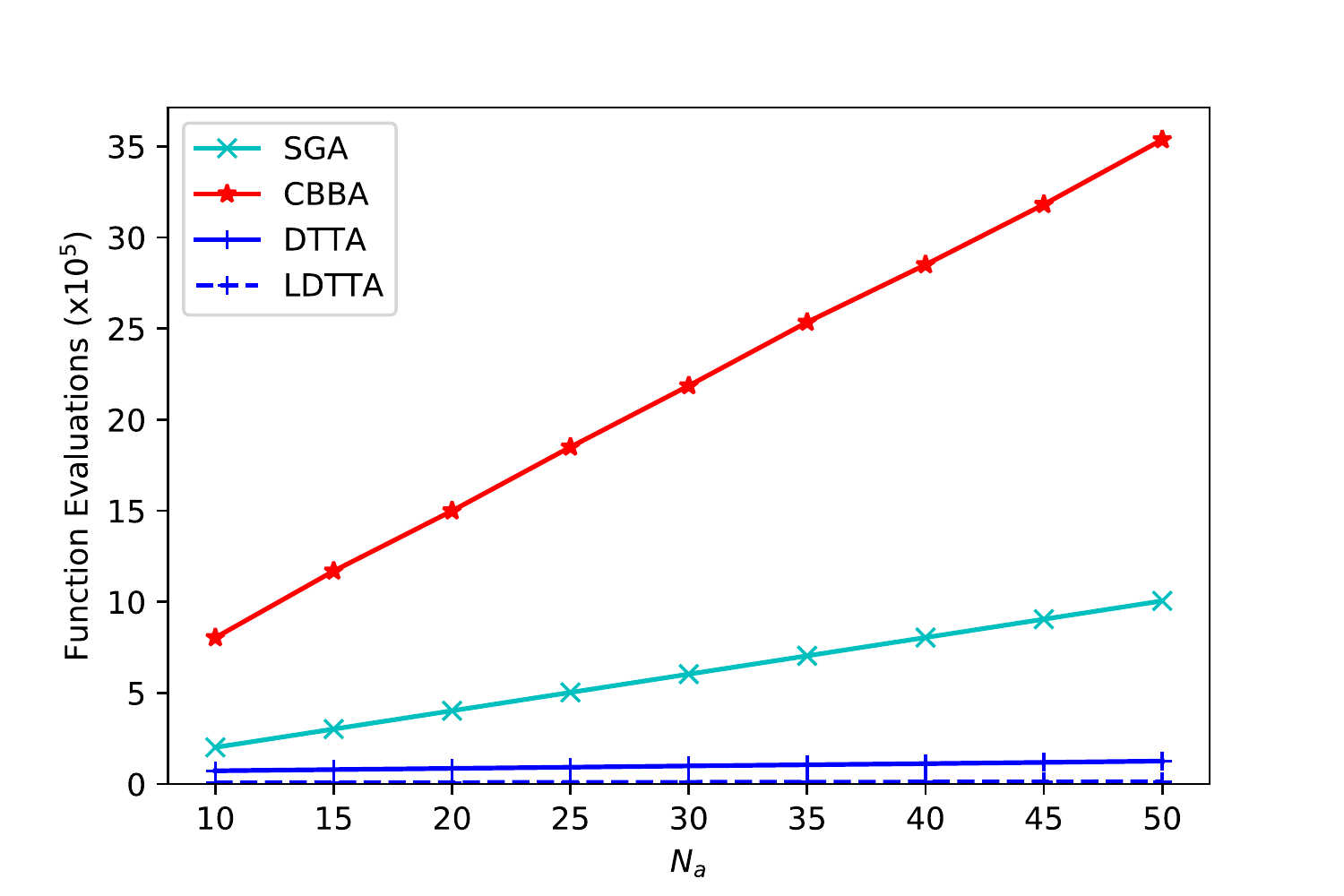}} \hfill
\vskip -10pt
\subfloat[Consensus Steps]{\includegraphics[width=0.46\textwidth]{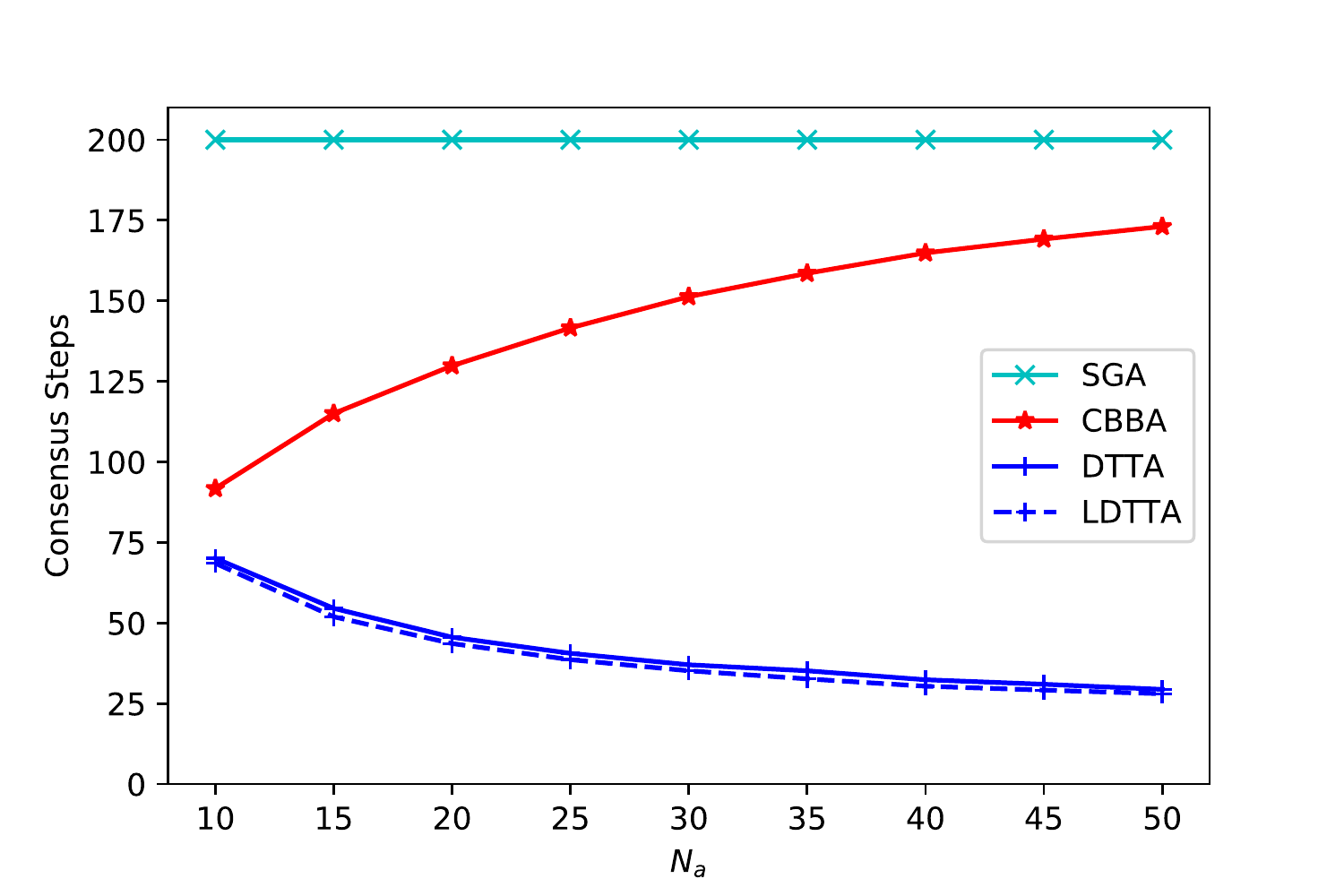}}\hfill
\subfloat[Running Time]{\includegraphics[width=0.46\textwidth]{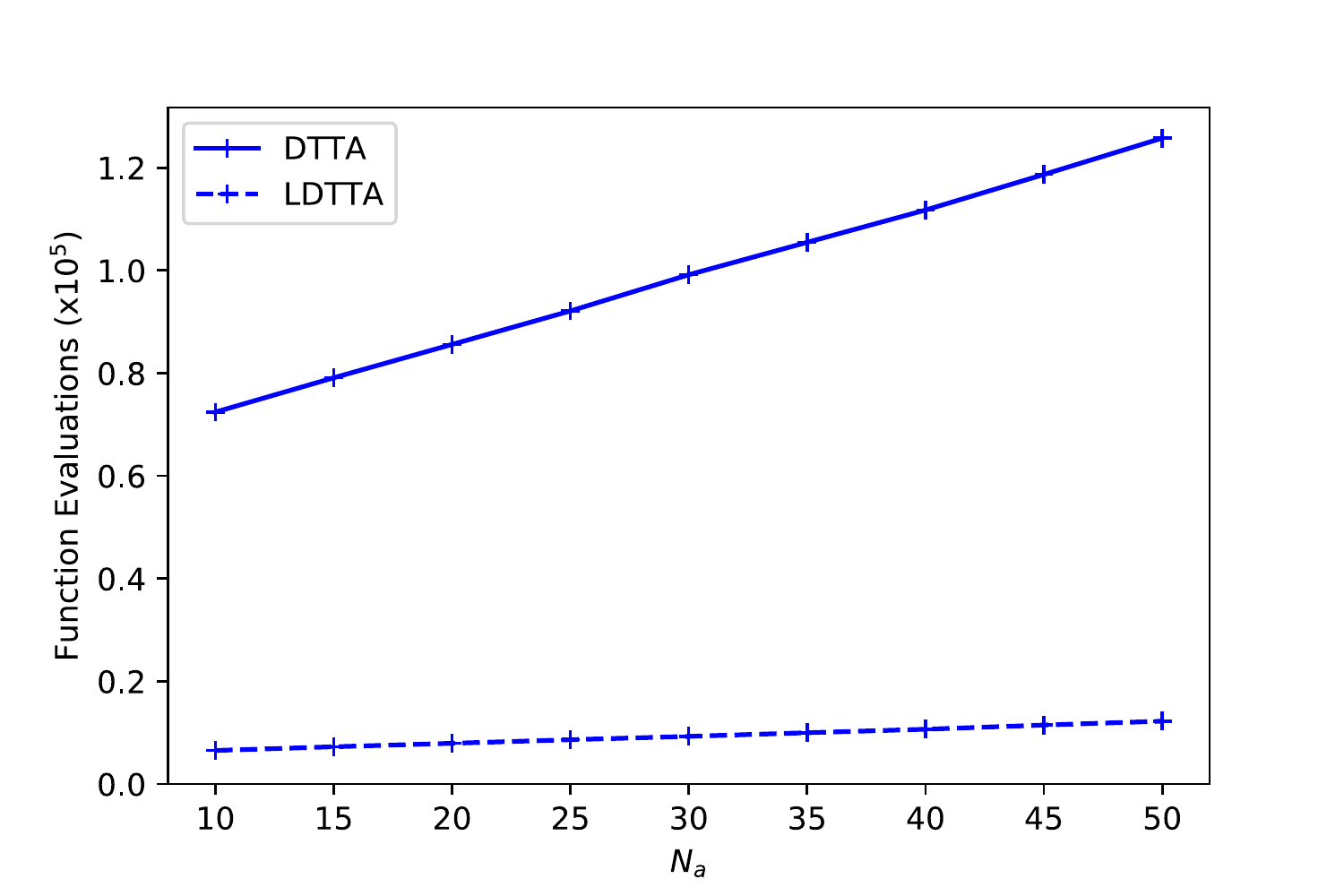}}\hfill
\vskip -10pt
\subfloat[Ratio Comparison $N_a=10$]{\includegraphics[width=0.46\textwidth]{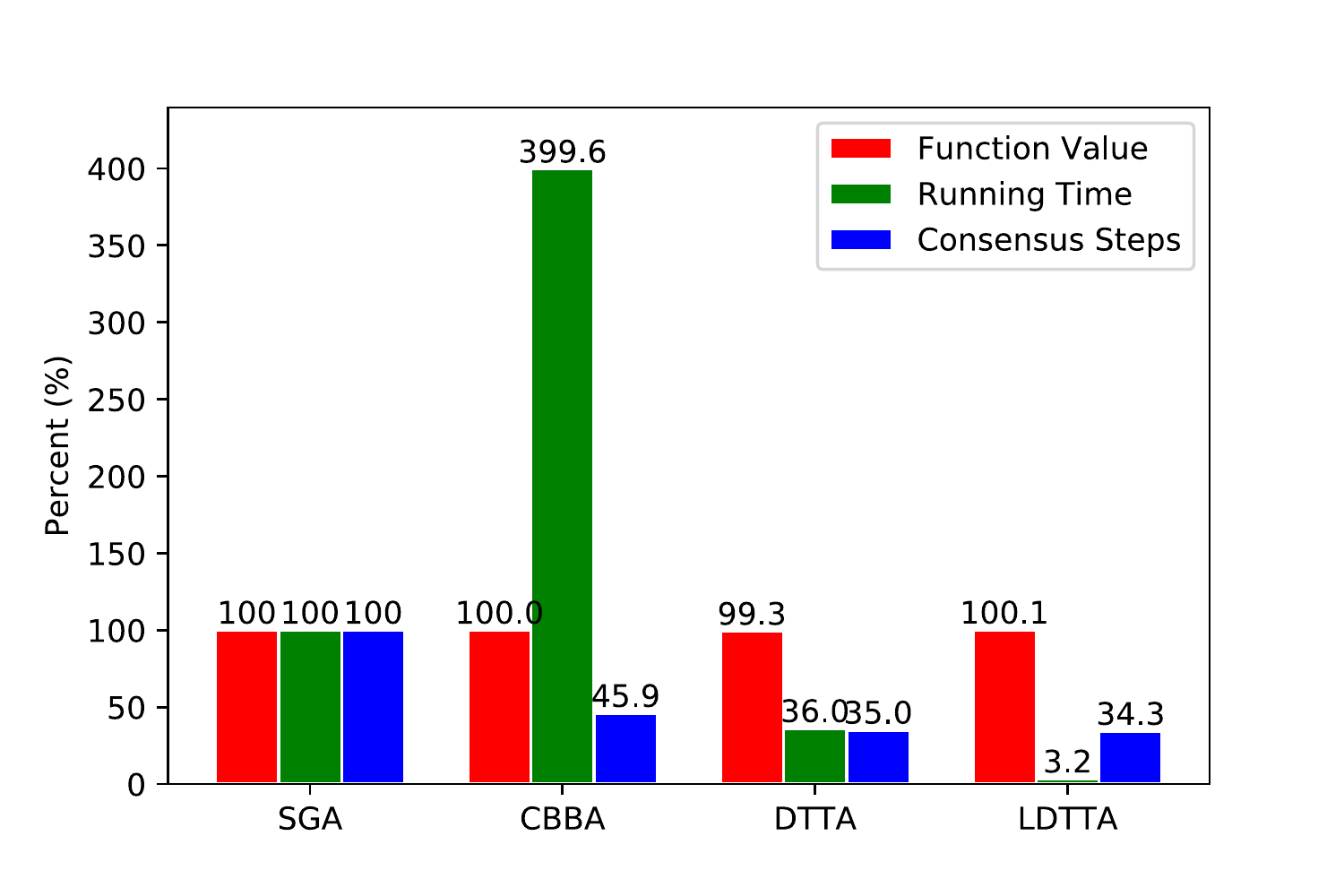}}\hfill
\subfloat[Ratio Comparison $N_a=50$]{\includegraphics[width=0.46\textwidth]{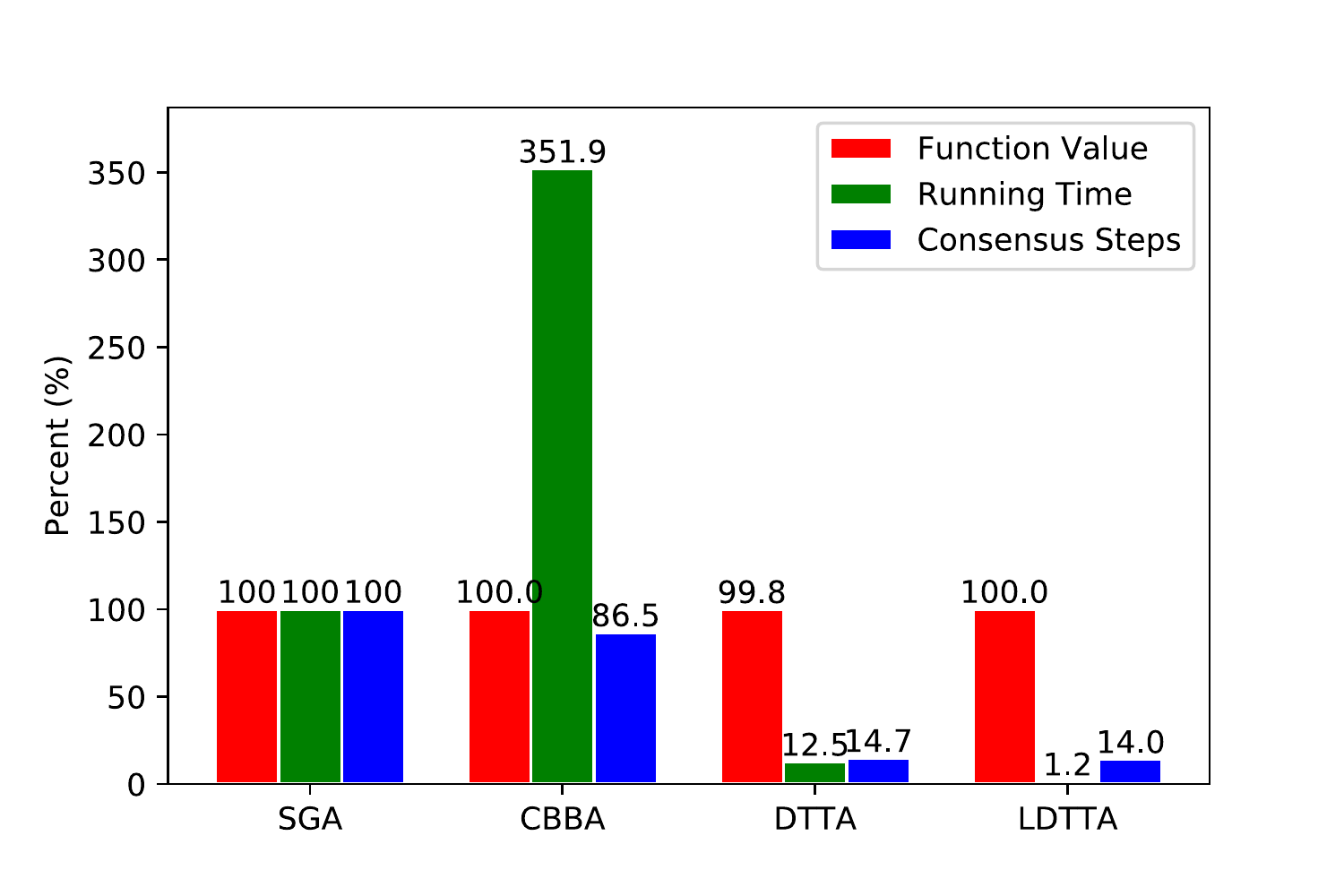}}
\caption{Performance comparison of the proposed algorithms and benchmark algorithms}
\label{fig: monotone comparison}
\end{figure*}
 

The performances of DTTA and LDTTA are compared with benchmark task allocation algorithms SGA and CBBA in Fig. \ref{fig: monotone comparison}. The advantage of CBBA is that it can reduce consensus steps with the help of building task bundles while achieving the same solution quality compared with SGA. Note that, the number of consensus steps required by SGA is equal to the number of tasks, i.e., 200 in this case. As shown in Fig. \ref{fig: monotone comparison} (c), (e), and (f), CBBA consumes 45.9\% of consensus steps of SGA when there are 10 UAVs. The percentage goes up to 86.5\% as the number of UAVs increases to 50. As mentioned in \cite{choi2009consensus}, agents need to continually rebuild their bundles to solve the conflict during the task allocation process. The issue with the bundle rebuilding is that when there are a large number of tasks, evaluating the objective function is quite time-consuming. This is the reason why CBBA performs poorly in terms of running time, as shown in Fig. \ref{fig: monotone comparison} (b), (e), and (f). By contrast, DTTA and LDTTA perform much better in terms of both consensus steps and running time while obtaining almost the same function values compared with SGA and CBBA. As shown in Fig. \ref{fig: monotone comparison} (f), when $N_a=50$, LDTTA consumes only 1.2\% of running time and 14.0\% of consensus steps with SGA as a baseline. Fig. \ref{fig: monotone comparison} (c) indicates that as the number of UAVs increases, unlike CBBA, the numbers of consensus steps for DTTA and LDTTA decrease because more tasks can be allocated within one consensus step when there are more UAVs. This is an extremely beneficial feature in the large-scale task allocation problems where scalability is one of the key concerns. 

\begin{figure*}[ht]
\subfloat[Function Value]{\includegraphics[width=0.46\linewidth]{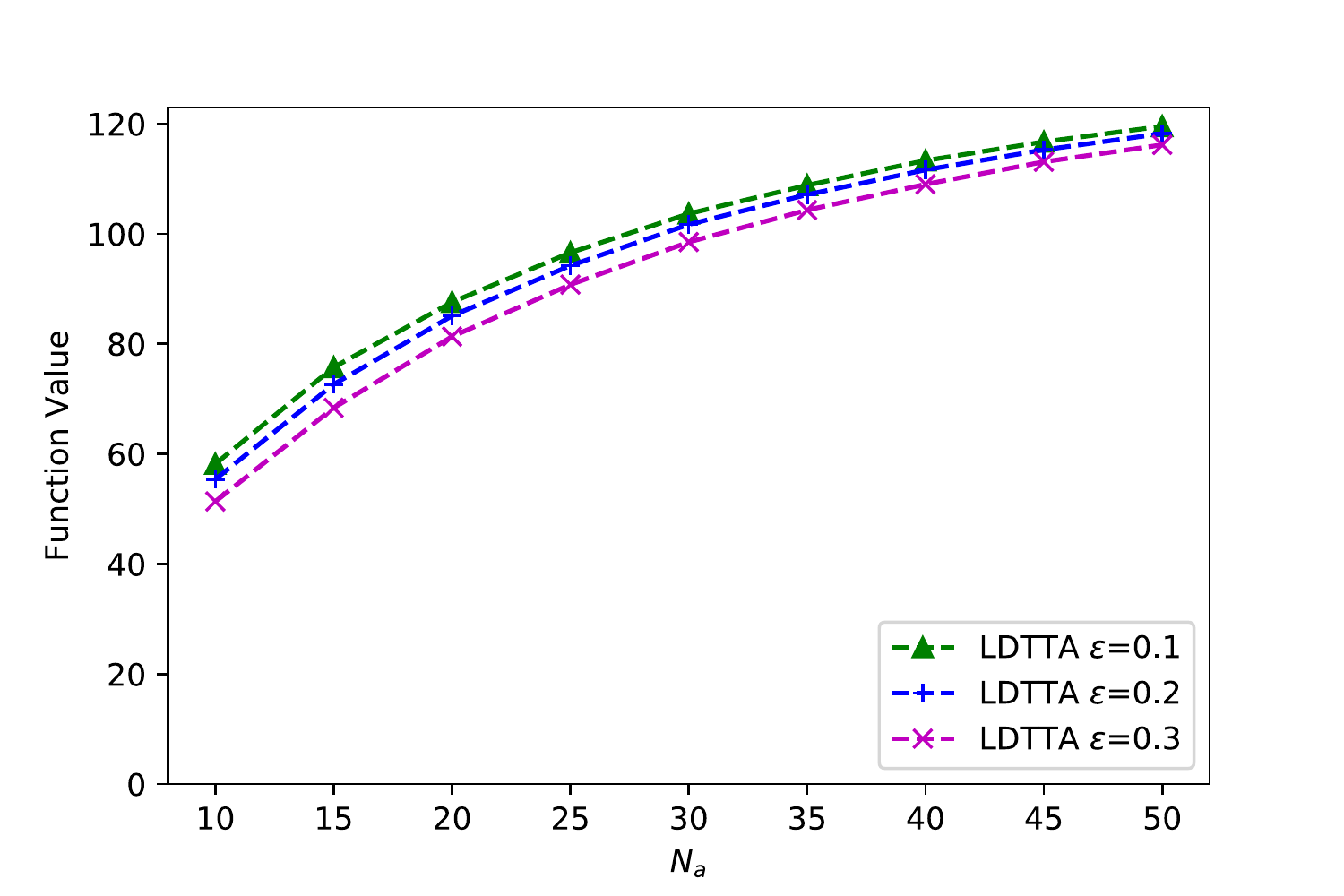}}\hfill
\subfloat[Running Time]{\includegraphics[width=0.46\textwidth]{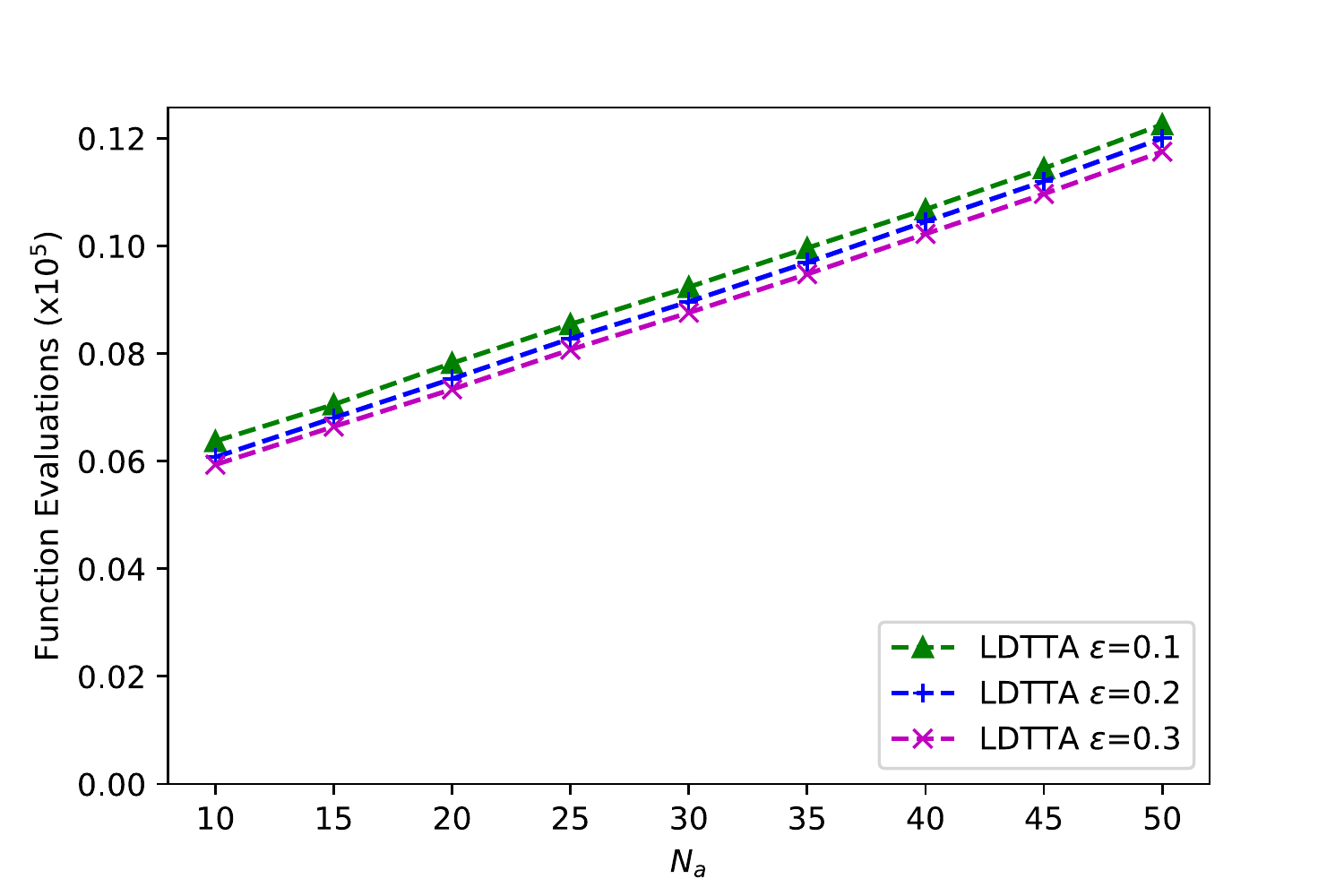}} \hfill
\vskip -10pt
\subfloat[Consensus Steps]{\includegraphics[width=0.46\textwidth]{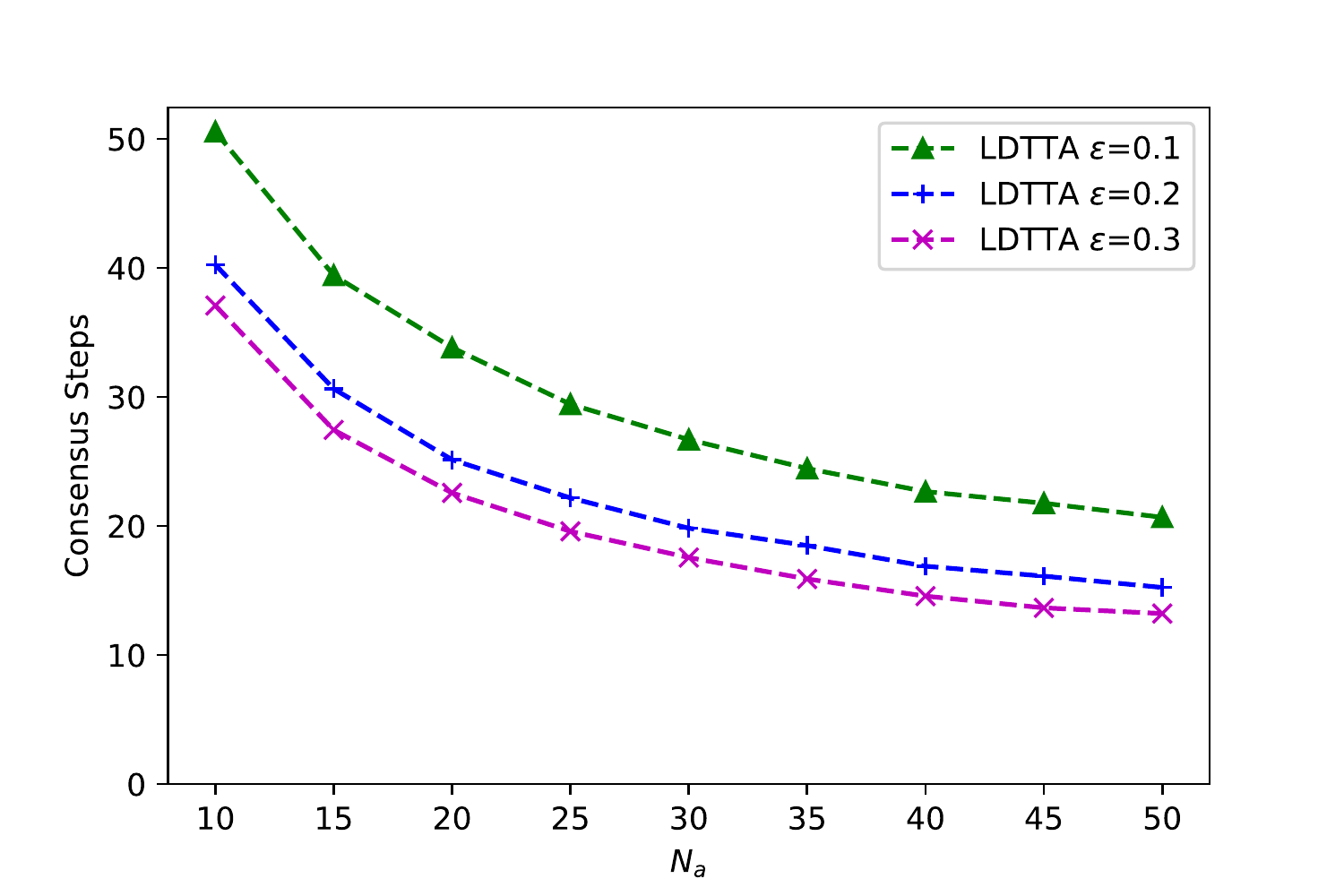}}\hfill
\subfloat[Ratio Comparison $N_a=30$]{\includegraphics[width=0.46\textwidth]{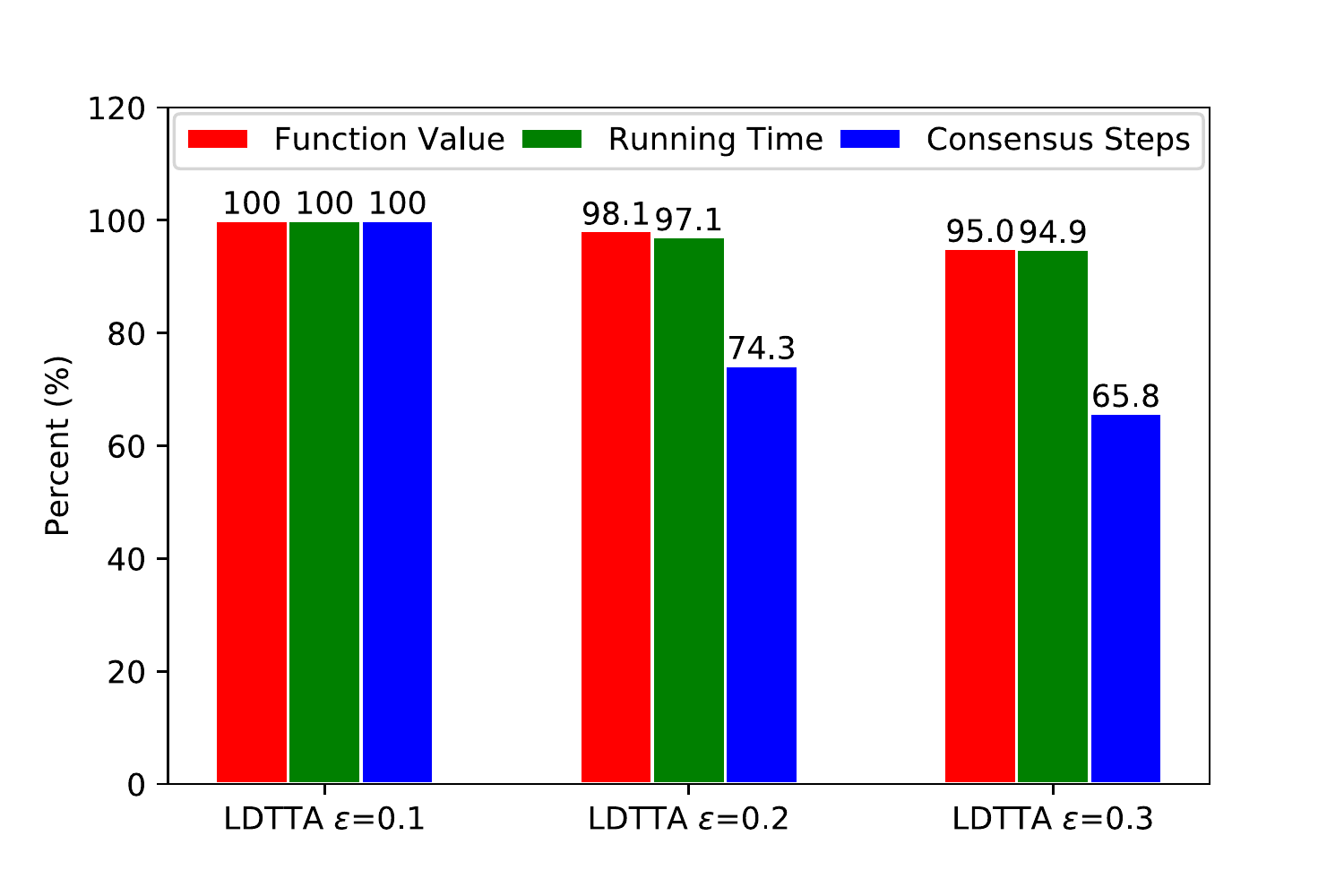}}
\caption{Trade-off of function value, running time and consensus steps}
\label{fig: TradeOff}
\end{figure*}

%

According to Fig. \ref{fig: monotone comparison} (d), (e), and (f), LDTTA consumes much less running time and slightly fewer consensus steps with slightly higher function values compared with DTTA under the same $\epsilon$. In LDTTA, the remaining tasks are sorted in descending order according to their marginal values. On average, LDTTA is more likely to find a qualified task with a higher marginal value, thereby achieving better solution quality than DTTA does. Meanwhile, since UAVs and tasks are heterogeneous and located at different positions, UAVs are less likely to have a conflict with each other using LDTTA than using DTTA. Thus, LDTTA consumes fewer consensus steps than DTTA. Overall, LDTTA has the best performance among all these algorithms. 

Fig. \ref{fig: TradeOff} demonstrates the trade-off performance of LDTTA where the threshold decreasing parameter $\epsilon$ is set as 0.1, 0.2, and 0.3 respectively. The results show that as $\epsilon$ increases, the function value, running time, and consensus steps decrease, which verifies the trade-off. Fig. \ref{fig: TradeOff} (d) compares the ratio of function value, running time, and consensus steps with $\epsilon=0.1$ as a baseline. It indicates that as the value of $\epsilon$ increases from 0.1 to 0.3, the ratios of function value and running time decrease slowly at a similar pace, while the ratio of consensus steps drops rapidly.


\section{Conclusions}
\label{sec: Conlusions}

This paper proposed two efficient decentralised algorithms, i.e., Decreasing Threshold Task Allocation (DTTA) and an upgraded version Lazy Decreasing Threshold Task Allocation (LDTTA), for Multi-Robot Systems (MRS) in large-scale task allocation problems. DTTA and LDTTA enabled parallel allocation with the help of a decreasing threshold hence releasing the computational and communicating burden for MRS. The performance of the proposed algorithms was analysed theoretically and verified through Monte-Carlo simulations of a multi-target surveillance mission using a group of heterogeneous UAVs. Simulation results indicated that the proposed algorithms consumed much less running time and consensus steps while achieving almost the same objective function values compared with the benchmark task allocation algorithms. The results of the proposed algorithms are expected instrumental in the large-scale physical applications where MRS usually suffer from the computational and communicating burden.

One future work would be further analysing the theoretical performance of the proposed algorithms by combining the curvature of the submodular objective function to provide a tighter optimality bound of the solution quality. Another future work would be verifying the performance of the proposed algorithms through physical experiments of a surveillance mission in a large area using multiple UAVs.


\bibliographystyle{plain}
\bibliography{DTTA_autart}

\appendix
\section{Analysis of the Objective Function}

In the appendix, we prove that the objective function Eqn. (\ref{eqn: f_final}) is monotonically increasing, non-negative, and submodular. We start the proof from analysing Eqn. (\ref{eqn: f_a}). 
    
For a random UAV $a$, assume that there are two task selection sets $\mathcal{T}_1$ and $\mathcal{T}_2$ satisfying $\mathcal{T}_1 \subseteq \mathcal{T}_2 \subseteq \mathcal{T}$. A new task $j \in \mathcal{T} \backslash \mathcal{T}_2$ is going to be considered by this UAV. $\mathbf{p}_1^j$ represents the path from initial position of UAV $a$ to the task $j$ along the task set $\mathcal{T}_1$ and $\mathbf{p}_2^j$ is defined in the same way with $\mathcal{T}_2$.
    
\begin{lemma}
\label{thm: monotone and non-negative}
The objective function Eqn. (\ref{eqn: f_a}) is non-negative and increasing.
\end{lemma}
\begin{pf}
By definition, if $\mathcal{T}_a = \emptyset$, then $f_a(\mathcal{T}_a)=0$ and these four factors $m_{aj}, v_j, \lambda_d, \lambda_n \in (0,1]$ are positive. Then the marginal value of task $j$ is always positive, i.e.
\begin{equation*}
m_{aj} v_j \lambda_d^{\tau(\mathbf{p}_a^j)} \lambda_n^{\sigma(\mathbf{p}_a^j)} > 0.
\end{equation*}
Eqn. (\ref{eqn: f_a}) is the sum of all marginal values, hence it is non-negative and increasing.
\end{pf}

\begin{lemma}
\label{thm:submodular}
The objective function Eqn. (\ref{eqn: f_a}) is submodular.
\end{lemma}

\begin{pf}
Since $\mathcal{T}_1 \subseteq \mathcal{T}_2$, the length of the path $\mathbf{p}_1^j$ is no greater than that of the path $\mathbf{p}_2^j$, and the number of tasks along the path $\mathbf{p}_1^j$ is no greater than that of the path $\mathbf{p}_2^j$, i.e.
\begin{equation}
\label{eqn: path length and number of tasks}
0<\tau(\mathbf{p}_1^j) \leq \tau(\mathbf{p}_2^j), ~ \\
0<\sigma(\mathbf{p}_1^j) \leq \sigma(\mathbf{p}_2^j).
\end{equation}
The marginal value of task $j$ given $\mathcal{T}_1$ is
\begin{equation*}
\Delta f_a(j|\mathcal{T}_1) = m_{aj} v_j \lambda_d^{\tau(\mathbf{p}_1^j)} \lambda_n^{\sigma(\mathbf{p}_1^j)}.
\end{equation*}
The marginal value of task $j$ given $\mathcal{T}_2$ is
\begin{equation*}
\Delta f_a(j|\mathcal{T}_2) = m_{aj} v_j \lambda_d^{\tau(\mathbf{p}_2^j)} \lambda_n^{\sigma(\mathbf{p}_2^j)}.
\end{equation*}
According to Eqn. (\ref{eqn: path length and number of tasks}) and since $\lambda_d, \lambda_n \in (0,1]$, we have
\begin{equation*}
\Delta f_a(j|\mathcal{T}_1) \geq \Delta f_a(j|\mathcal{T}_2) .
\end{equation*}
Therefore, Eqn. (\ref{eqn: f_a}) is submodular.
\end{pf}

The task sets for all UAVs are disjoint since one task can be allocated to no more than one UAV. According to the property of submodularity, the sum of submodular functions is submodular. Therefore, the objective function Eqn. (\ref{eqn: f_final}) is monotonically increasing, non-negative, and submodular.


\end{document}